\documentclass[journal]{IEEEtran}
\usepackage{amsmath}
\usepackage{amsfonts}
\usepackage{amssymb}
\usepackage{amsthm}
\usepackage{graphicx}
\usepackage{dcolumn}
\usepackage{bm}
\usepackage{color}

\newcommand{\ket}[1]{\ensuremath{\left|{#1}\right\rangle}}
\newcommand{\bra}[1]{\ensuremath{\left\langle{#1}\right |}}

\newtheorem{theorem}{Theorem}
\newtheorem{lemma}{Lemma}
\newtheorem{corollary}[theorem]{Corollary}

\theoremstyle{definition}

\newtheorem{problem}{Problem}
\newtheorem{definition}{Definition}

\theoremstyle{remark}
\newtheorem{remark}[theorem]{Remark}

\begin{document}
%
\title{Ancilla-Assisted Discrimination of Quantum Gates}
%
%
%

\author{Jianxin~Chen~\IEEEmembership{}
        and~Mingsheng~Ying~\IEEEmembership{}
\thanks{J. Chen and M. Ying are with the State Key Laboratory of Intelligent Technology and
Systems, Tsinghua National Laboratory for Information Science and
Technology, Department of Computer Science and Technology, Tsinghua
University, Beijing 100084, China. e-mail: chenkenshin@gmail.com (J. Chen), yingmsh@mail.tsinghua.edu.cn (M. Ying)}
}

\maketitle

\begin{abstract}
The intrinsic idea of superdense coding is to find as many gates as
possible such that they can be perfectly discriminated. In this
paper, we consider a new scheme of discrimination of quantum gates,
called ancilla-assisted discrimination, in which a set of quantum
gates on a $d-$dimensional system are perfectly discriminated with
assistance from an $r-$dimensional ancilla system. The main
contribution of the present paper is two-fold: (1) The number of
quantum gates that can be discriminated in this scheme is evaluated.
We prove that any $rd+1$ quantum gates cannot be perfectly
discriminated with assistance from the ancilla, and there exist $rd$
quantum gates which can be perfectly discriminated with assistance
from the ancilla. (2) The dimensionality of the minimal ancilla
system is estimated. We prove that there exists a constant positive number
$c$ such that for any $k\leq cr$ quantum
gates, if they are $d$-assisted discriminable, then they are also
$r$-assisted discriminable, and there are $c^{\prime}r\textrm{
}(c^{\prime}>c)$ different quantum gates which can be discriminated
with a $d-$dimensional ancilla, but they cannot be discriminated if
the ancilla is reduced to an $r-$dimensional system. Thus, the order
$O(r)$ of the number of quantum gates that can be discriminated with
assistance from an $r-$dimensional ancilla is optimal. The results
reported in this paper represent a preliminary step toward
understanding the role ancilla system plays in discrimination of
quantum gates as well as the power and limit of superdense coding.
\end{abstract}

\begin{IEEEkeywords}
superdense coding, discrimination, quantum gates, ancilla system, algebraic geometry.
\end{IEEEkeywords}

\section{Introduction}\label{section:intro}
\IEEEPARstart{H}{ow} to determine the oracle from many known candidates if the
investigation into internal structures of oracle is not allowed?
This is one of the central questions in theoretical computer science
named as oracle identification problem or sometimes circuit verification problem.

In quantum information theory, a special case of the oracle
identification problems is the discrimination of quantum states \cite{Yuen75}.
This is a fundamental task in quantum information theory since
information is encoded in states and after accomplishing some
quantum information tasks, information encoded in the states has to
be read out, that means the final state of the system has to be
determined. It is well known that perfect discrimination is possible
only for orthogonal quantum states~\cite{Bennett92}.  Thus,
so-called unambiguous discrimination strategy is introduced for
quantum states which are not necessarily mutually
orthogonal~\cite{Ivanovic87, Dieks88, Peres88, Jaeger95, Chefles98,
Chefles01,  Eldar03A, Zhang01, Chen01, Feng02, Feng04,
Jafarizadeh08}. In contrast to perfect discrimination where one can
always identify the state, unambiguous discrimination guarantees that except for an
inconclusive probability, one can always get the correct state with
zero error probability. It was shown that a set of quantum pure
states are unambiguously discriminable if and only if they are
linearly independent~\cite{Chefles98}. Furthermore, many variations
of state discrimination problem such as local discrimination,
discrimination of mixed states, discrimination of multipartite
quantum states, minimum-error discrimination and asymptotic discrimination were proposed and have
attracted much attention~\cite{Walgate00, Duan0702, Xin08,
Rudolph03,Takeoka03,Raynal03, Eldar01, Eldar03B, Eldar04, Schmid2008, Markham2008}. For a
comprehensive survey of state discrimination, we refer to~
\cite{Chefles00, Paris04}.

Another class of the oracle identification problems arisen in
quantum information theory is the discrimination of quantum gates
which is indeed closely related to the discrimination of quantum
states. The problem of discriminating oracle quantum gates can be
formally stated as follows. Suppose we are given an unknown quantum
gate, namely a unitary operator, which is secretely chosen from a
set $\{U_1,...,U_k\}$ of quantum gates acting on a quantum system whose
state space is a $d-$dimensional Hilbert space $\mathcal{H}_d$. The
aim is to decide which one is the given gate. The discrimination of
$U_1,...,U_k$ can be realized by transforming it to a corresponding
problem of discriminating quantum states, more precisely, by
choosing a suitable input state $|\psi\rangle\in\mathcal{H}_d$ and
then discriminating $U_1|\psi\rangle,...,U_k|\psi\rangle$.

Several variants of the problem of discriminating quantum operations have been considered in the literature. For example, unambiguous discrimination is introduced to the discrimination of
quantum operations~\cite{Chefles07, Wang06}. If we consider quantum
gates acting on many particles, discrimination using local
operations and classical communication is then proposed and has been
extensively studied~\cite{Zhou07, Duan07, Duan08}.

It should be remarked that, for perfect discrimination of quantum
gates, it was shown that any two different quantum gates can always
be perfectly discriminated if multicopy is allowed, which is quite
different from perfect discrimination of quantum
states, where infinite copies are required in order to
attain perfect discrimination~\cite{Acin01}. Thus the multicopy strategy would greatly extends the ability to discriminate quantum gates.

In this paper, we consider another possibility of improving ability
of discriminating quantum gates, namely discrimination with an
ancilla system. If ancilla system is allowed, then the ability of discriminating quantum gates acting on $d$-dimensional Hilbert space should be improved. However, any reasonable ancilla system in practice generally cannot possibly extend to large scale. Thus we assume the dimension of ancilla system is bounded by some $r\leq d$. In this situation, we say these quantum gates can be $r$-assisted discriminated.

The purpose of the present paper is to address the following two questions and to explore the role ancilla system plays in discriminating among quantum oracle gates.

\begin{problem}\label{Problem:1}[\textbf{Maximal Discrimination Problem}]
How many quantum gates acting on $d$-dimensional Hilbert space at most can be perfectly discriminated with assistance from an $r$-dimensional ancilla?
\end{problem}

\begin{problem}\label{Problem:2}[\textbf{r-Reduced Discrimination Problem}]
What is the maximum number $N$ such that for any $N$ quantum gates
acting on $d$-dimensional Hilbert spaces, if they can be perfectly
discriminated with some large ancilla system, then they can still be
discriminated when the ancilla system is reduced to $r$-dimensional.
\end{problem}

Ancilla-assisted discrimination of quantum gates is a fundamental
problem in quantum information theory. Its importance is witnessed
by the simple observation that the intrinsic idea of superdense
coding is to find as many gates as possible such that they can be
perfectly discriminated by some state~\cite{Bennett92, mozes05}. To
be more precise, let's recall the protocol for standard superdense
coding. Suppose there are two parties, known as Alice and Bob, who
are far away from one another. Alice possesses a $p$-qubit system
$\mathcal{A}$ and Bob holds a $q$-qubit system $\mathcal{B}$. Their
goal is to transmit as much information as possible. Then Alice can
send $k$ bits of classical information to Bob using only $p$ qubits
by following procedure if there exist ${2^k}$ quantum gates
$\{U_i\}_{i=1}^{2^k}$ acting on $\mathcal{A}$ which can be
$2^q$-assisted discriminable. Let Alice and Bob initially share a
state $\ket{\psi}_{\mathcal{A}\mathcal{B}}$ which satisfies
\begin{displaymath}
\bra{\psi}_{\mathcal{A}\mathcal{B}}{(U_i\otimes I_{\mathcal{B}})}^{\dagger}(U_j\otimes I_{\mathcal{B}})\ket{\psi}_{\mathcal{A}\mathcal{B}}=0
\end{displaymath}
for all $1\leq i\neq j\leq 2^k$. The existence of such $\ket{\psi}$ is equivalent to $2^q$-assisted discriminability of these quantum gates. If Alice wishes to send the bit string '$i_1i_2\cdots i_k$' to Bob, she applies the quantum gate $U_{i_1i_2\cdots i_k}$ to her qubits, and then sends her qubits to Bob, giving Bob possession of the whole system. Notice that
\begin{displaymath}
\bra{\psi}_{\mathcal{A}\mathcal{B}}{(U_i\otimes I_{\mathcal{B}})}^{\dagger}(U_j\otimes I_{\mathcal{B}})\ket{\psi}_{\mathcal{A}\mathcal{B}}=0
\end{displaymath}
for all $1\leq i\neq j\leq 2^k$
is equivalent to $\{(U_i\otimes I_{\mathcal{B}})\ket{\psi}_{\mathcal{A}\mathcal{B}}\}_{i=1}^{2^k}$ form a subset of an orthogonal basis.
Thus Bob can do a measurement in this basis, and then determine which of the bit strings Alice sent. Thus the natural strategy of optimal superdense coding scheme is to find out as many unitary gates as possible, which can be perfectly discriminated with assistance from $\mathcal{B}$.

The paper is structured as follows. After introducing our notations
and giving some background on discrimination of quantum gates in
section~\ref{section:intro}, we give a straightforward and complete
solution to Problem~\ref{Problem:1} in section~\ref{section:1}.
Section ~\ref{section:r} is devoted to answer
Problem~\ref{Problem:2}. This section is divided into two
subsections. In the first subsection, we present an upper bound for
Problem~\ref{Problem:2}, and a lower bound for this problem is given
in the second subsection. To prove the theorem concerning the lower
bound, some basic ideas from algebraic geometry are required. For
convenience of the reader, the necessary concepts and theorems in
algebraic geometry are included in the second subsection. Finally,
we conclude and present a number of open problems.

Through this paper, we focus on perfectly discrimination strategy. And for simplicity, sometimes we may use the word ``discrimination'' directly instead of ``perfectly discrimination'' without explicit explanation.

\section{Maximal Discrimination Problem}\label{section:1}

In this section, we will answer Problem~\ref{Problem:1}. We first
give a formal definition of discriminability of quantum gates.

\begin{definition}
Suppose a $d$-dimensional Hilbert space $\mathcal{H}_d$ is given,
consider a set $\mathcal{U}=\{U_i\}_{i=1}^k$ of quantum gates acting
on $\mathcal{H}_d$.
\begin{enumerate}
\item We say $\mathcal{U}$
is (or $U_1,...,U_k$ are) discriminable if it (or they) will output a set of
orthogonal states for some given initial input state; that is, there
exists $\ket{\psi}\in \mathcal{H}_d$ such that
$\bra{\psi}{U_i}^{\dagger}{U_j}\ket{\psi}=0$ for all $1\leq i\neq
j\leq k$.
\item We say $\mathcal{U}$ is (or $U_1,...,U_k$ are)
$r$-assisted discriminable if it is (or they are) discriminable with an
$r$-dimensional ancilla system $\mathcal{H}_R$; that is, there
exists a pure state $\ket{\psi}\in \mathcal{H}_d\otimes
\mathcal{H}_R$ such that $\bra{\psi}{({U_i}\otimes
I_R)}^{\dagger}({U_j}\otimes I_R)\ket{\psi}=0$ for all $1\leq i\neq
j\leq k$.\end{enumerate}
\end{definition}

The following two technical lemmas will be needed in the proof of
our main theorem in this section.

\begin{lemma}\label{lemma:density}
$k$ quantum gates $\{U_i\}_{i=1}^k$ are $r$-assisted discriminable
if and only if there exists a density operator $\rho$ with rank
$\leq r$ such that
\begin{displaymath}
tr(U_i^{\dagger}U_j\rho)=0\textrm{ }for \textrm{ }all \textrm{
}1\leq i\neq j\leq k.
\end{displaymath}
\end{lemma}
\IEEEproof{Assume $\{U_i\}_{i=1}^k$ are $r$-assisted discriminable. Then there
exists a state $\ket{\psi}\in \mathcal{H}_d\otimes \mathcal{H}_R$
such that $$\bra{\psi}{(U_i\otimes I_R)}^{\dagger}(U_j\otimes
I_R)\ket{\psi}=0$$ for all $1\leq i\neq j\leq k$. We noticed that
any $\ket{\psi}\in \mathcal{H}_d\otimes \mathcal{H}_R$ can be
expressed as $\sum\limits_{l=0}^{r-1}\ket{\psi_l}\ket{l}_{R}$, where
$\{\ket{l}_{R}\}_{l=0}^{r-1}$ form an orthogonal basis of the
ancilla system, and $\ket{\psi_l}\in \mathcal{H}_d$ are unnormalized
states for all $0\leq l \leq r-1$. It follows immediately that
\begin{eqnarray}
\sum\limits_{l=0}^{r-1}&&\bra{\psi_l}{\bra{l}}_R(U_i\otimes
I_R)^{\dagger}(U_j\otimes I_R)\sum\limits_{l^{\prime}=0}^{r-1}
\ket{\psi_{l^{\prime}}}{\ket{l^{\prime}}}_R \nonumber\\
&&=\sum\limits_{l=0}^{r-1}\bra{{\psi_l}}U_i^{\dagger}U_j{\ket{\psi_l}}\nonumber\\
&&=tr(U_i^{\dagger}U_j\sum\limits_{l=0}^{r-1}\ket{\psi_l}\bra{\psi_l}),\nonumber
\end{eqnarray}
So, $\rho=\sum\limits_{l=0}^{r-1}\ket{\psi_l}\bra{\psi_l}$ is the
required density operator and $rank(\rho)\leq r$.

Conversely, if a density operator $\rho$ with rank $\leq r$ is
given, with $tr(U_i^{\dagger}U_j\rho)=0$ for all $1\leq i\neq j\leq
k$, then we can use the spectrum decomposition to find unnormalized
states $\{\ket{\psi_l}\}_{l=0}^{r-1}$ satisfying
$\rho=\sum\limits_{l=0}^{r-1}\ket{\psi_l}\bra{\psi_l}$, and further
construct the corresponding
$$\ket{\psi}=\sum\limits_{l=0}^{r-1}\ket{\psi_l}\ket{l}_R\in
\mathcal{H}_d\otimes \mathcal{H}_R.$$ It is obvious that such
$\ket{\psi}$ satisfies $\bra{\psi}{U_i}^{\dagger}{U_j}\ket{\psi}=0$
for all $1\leq i\neq j\leq k$ which means these quantum gates are
$r$-assisted discriminable. Thus, we have proved our lemma and sometimes we say that $\{U_i\}_{i=1}^k$ are perfectly discriminated by $\rho$.
}

Recall that a set $\{M_i\}_{i=1}^k$ of matrices are said to be
mutually orthogonal if $tr(M_i^{\dagger}M_j)=0$ for all $1\leq i\neq
j\leq k$.

\begin{lemma}\label{lemma:orthonal} $k$ quantum gates $\{U_i\}_{i=1}^k$ acting on a $d$-dimensional
Hilbert space are $r$-assisted discriminable if and only if there
exists a matrix $S$ with rank $\leq r$ such that $\{U_iS\}_{i=1}^k$
are mutually orthogonal; in this case we say that $\{U_i\}_{i=1}^k$
are discriminated by $S$.
\end{lemma}
\IEEEproof{From Lemma~\ref{lemma:density}, $\{U_i\}_{i=1}^k$ are
$r$-assisted discriminable if and only if there exists a density
operator $\rho$ with rank $\leq r$ such that
$tr(U_i^{\dagger}U_j\rho)=0$ for all $1\leq i\neq j\leq k$. Note
that the density operator $\rho$ can be written as $SS^{\dagger}$
for some matrix $S$, and $tr(U_i^{\dagger}U_j\rho)=0$ means $U_iS$
is orthogonal to $U_jS$. Thus $\{U_iS\}_{i=1}^k$ are mutually
orthogonal. The converse is also straightforward, and we omit the
proof here. }

Now we are able to present the main result of this section which
give a complete solution to Problem 1.

\begin{theorem}\label{theorem:upper}If $k$ quantum gates $\{U_i\}_{i=1}^k$ acting on $d$-dimensional Hilbert
space are $r$-assisted discriminable, then $k\leq rd$, and the upper
bound $rd$ can be achieved provided $r\leq d$.
\end{theorem}
\IEEEproof{Suppose $\{U_i\}_{i=1}^k$ are
$r$-assisted discriminable. Then from Lemma~\ref{lemma:orthonal} we
know that there exists a matrix $S$ with rank $\leq r$ satisfying
mutual orthogonality of $\{U_iS\}_{i=1}^k$. We can write $S$ in the
following way: $$S= \sum\limits_{k=1}^r\ket{p_k}\bra{q_k}.$$ Then
$U_iS$ lies in the subspace $H_d\otimes span(\{\bra{q_j}\}_{j=1}^r)$
whose dimension is at most $rd$. This together with the fact that
$\{U_i\}_{i=1}^k$ are mutually orthogonal implies $k\leq rd$.

To illustrate our observation, we introduce several elementary
unitary gates first. Suppose $\{\ket{k}\}_{k=1}^d$ is the computational basis of
a qudit space $\mathcal{H}_d$. We define $X$ by
$X\ket{k}=\ket{k+1},$ where the addition is modulo $d$, and $Z$ by
$Z\ket{k}=\omega^k\ket{k},$ where $\omega=e^{-\frac{2\pi i}{d}}$ is
the $d$th unity root. Then the set of the generalized Pauli matrices
in the qudit system  is $\{X^pZ^q:p,q=0,\cdots,d-1\}$.

Now we assume that $r\leq d$. For the simplest case of $r=1$, we
choose $d$ gates: $\{X^i\}_{i=0}^{d-1}$.
Obviously, they can be discriminated by $diag\{1,0,\cdots, 0\}$. For
the case of $r\geq 2$, let $\omega$ be the unity root of order $r$,
and let
$$W=diag\{1, \omega, \omega^2, \cdots, \omega^{r-1}, 1, \cdots,
1\}.$$ We choose $rd$ quantum gates: $\{X^iW^j\}$, where $0\leq
i\leq r-1,\textrm{ } 0\leq j\leq d-1$. It is easy to see that they
can be discriminated by $\rho=diag\{\frac{1}{r},\cdots,\frac{1}{r},0,\cdots,0\}$ whose first
$r$ entries in the principle diagonal are all $\frac{1}{r}$ and others are all
zero. Thus, we complete the proof by Lemma~\ref{lemma:density}.
}

As a simple application of the above theorem, we return back to the
superdense coding scenario considered in the Introduction. With
Theorem~\ref{theorem:upper}, we have
\begin{displaymath}
2^k\leq dim(\mathcal{A}) dim(\mathcal{B})=2^p2^q=2^{p+q}.
\end{displaymath}
which follows $k\leq p+q$. Thus if Alice possesses $p$ qubits and Bob possess $q$ qubits, then Alice can transmit at most $p+q$ bits of classical information to Bob using her $p$ qubits of quantum information, this bound can be achieved provided $q\leq p$.

\section{r-Reduced Discrimination Problem}\label{section:r}

As pointed in the Introduction, this section is entirely devoted to
address Problem 2. Suppose a set $\mathcal{U}$ of quantum gates
acting on $d$-dimensional Hilbert space can be perfectly
discriminated with assistance from an ancilla system. In the real
world, sometimes the large ancilla system is not possible in
practice and what we can deal with is a small ancilla system. In
this section, we assume some quantum gates can be perfectly
discriminated with assistance from a large ancilla system, and our
goal is to reduce the ancilla such that these quantum gates still
can be perfectly discriminated with assistance from the reduced
ancilla system.

A simple application of the Schmidt decomposition indicates that the
quantum gates acting on $d$-dimensional Hilbert space can always be
perfectly discriminated with assistance from a $d$-dimensional
ancilla system; in other words, they are always $d$-assisted
discriminable. So, it is reasonable to rephrase Problem 2 as the
following more specific question:

\begin{problem}\label{Problem:3}[\textbf{r-Reduced Discrimination
Problem}] Given a positive integer $r<d$. What is the maximum number
$N$ satisfying that any $N$ $d$-assisted discriminable quantum gates
acting on $d$-dimensional Hilbert space will preserve their
discriminability when the ancilla system is reduced to dimension
$r$.
\end{problem}

We will answer this question by showing that, there exist $O(r)$
$d$-assisted discriminable quantum gates acting on $d$-dimensional
Hilbert space which cannot be $r$-assisted discriminated. On the
other hand, there exists some constant positive number $c$ such that for any $cr$
quantum gates, if they are perfectly discriminable with assistance
from some large ancilla system, then the discriminability remains
when the ancilla system is reduced to $r$-dimensional.

\subsection{Upper Bound of $r$-Reduced Discrimination problem}

In this subsection, we will show the upper bound for Problem 3 in
the following three steps.
\begin{enumerate}
 \item [1.] Starting from a simple observation at the ancilla-free discrimination.
 \item [2.] We will derive an upper bound $2\sqrt{rd}$ for general situation.
 \item [3.] If $r\leq \frac{d}{2}-1$, then our upper bound can be improved to $2r$.
\end{enumerate}

From above arguments it follows immediately that there always exist
$O(r)$ $d$-assisted discriminable quantum gates acting on
$d$-dimensional Hilbert space which cannot be $r$-assisted
discriminated.

Let's label the diagonals of a $d\times d$ matrix by integers $k$,
with $k$ increasing from lower left to upper right. So there are
$2d-1$ diagonals, and the $i$-th diagonal has $d-|d-x|$ entries. It
is easy to see that entries of $X^k$ are all 1's in the $(d-k)$-th
diagonal or $(2d-k)$-th diagonal, and all 0's in other diagonals.

\begin{lemma}\label{lemma:free}
There does not exist a state $\ket{\psi}\in \mathcal{H}_d$ such that
$\bra{\psi}XZ^k\ket{\psi}=0$ for all $0\leq k\leq d-1$ and
$\bra{\psi}Z^k\ket{\psi}=0$ for all $1\leq k\leq d-1$.
\end{lemma}
\IEEEproof{ If there exists a state $\ket{\psi}\in \mathcal{H}_d$ satisfies
the equations in the lemma, then it can be written in the
computational basis as
$\ket{\psi}=\sum\limits_{i=0}^{d-1}\lambda_i\ket{i}$, and
\begin{eqnarray}
\bra{\psi}XZ^k\ket{\psi}=0\nonumber
\iff&& (\sum\limits_{i=0}^{d-1}\lambda_i^{*}\bra{i})XZ^k(\sum\limits_{j=0}^{d-1}\lambda_j\ket{j})=0\nonumber\\
\iff&& \sum\limits_{i,j=0}^{d-1}\lambda_i^{*}\lambda_j\bra{i}XZ^k\ket{j}=0\nonumber\\
\iff&&
\sum\limits_{j=0}^{d-1}\omega^{kj}\lambda_{j+1}^{*}\lambda_j=0.\nonumber
\end{eqnarray}
Since $\bra{\psi}XZ^k\ket{\psi}=0$ for all $0\leq k\leq d-1$, we
have the following linear equations expressed in the matrix form:
\begin{displaymath}
\left(
\begin{array}{cccc}
1& 1 &\cdots& 1\\
1& \omega&\cdots& \omega^{d-1}\\
\vdots&\vdots&\ddots&\vdots\\
1&\omega^{d-1}&\cdots&\omega^{(d-1)(d-1)}
\end{array}
\right)
\left(
\begin{array}{c}
\lambda_1^{*}\lambda_0\\
\lambda_2^{*}\lambda_1\\
\vdots\\
\lambda_d^{*}\lambda_{d-1}
\end{array}
\right)=\mathcal{O},
\end{displaymath}
which implies $\lambda_{i+1}^{*}\lambda_i=0$ for all $1\leq i\leq
d$.

On the other hand, $\bra{\psi}Z^k\ket{\psi}=0$ with $1\leq k\leq
d-1$ can be rewritten as the following matrix equation:
\begin{displaymath}
\left(
\begin{array}{cccc}
1& \omega&\cdots& \omega^{d-1}\\
1& \omega^2&\cdots& \omega^{2(d-1)}\\
\vdots&\vdots&\ddots&\vdots\\
1&\omega^{d-1}&\cdots&\omega^{(d-1)(d-1)}
\end{array}
\right)
\left(
\begin{array}{c}
\lambda_0^{*}\lambda_0\\
\lambda_1^{*}\lambda_1\\
\vdots\\
\lambda_{d-1}^{*}\lambda_{d-1}
\end{array}
\right)=\mathcal{O}
\end{displaymath}
Thus we have $\vert
\lambda_0\vert=\cdots\vert\lambda_i\vert=\cdots\vert
\lambda_{d-1}\vert$. Combining this with the previous conclusion
$\lambda_{i+1}^{*}\lambda_i=0$ for all $1\leq i\leq d$, we obtain
$\ket{\psi}=0$.
}

\begin{lemma}
There exist  $O(\sqrt{d})$ quantum gates acting on a $d$-dimensional
Hilbert space which are $d$-assisted discriminable, but cannot be
perfectly discriminated if the ancilla system is not allowed.
\end{lemma}
\IEEEproof{
We choose the following quantum gates:
\begin{displaymath}
\begin{array}{cccc}
Z & Z^2 & \cdots & Z^{\lfloor \frac{p}{2}\rfloor}\\
I & Z^{p}&\cdots &Z^{\lceil \frac{d}{p}\rceil p}\\
X & XZ^p &\cdots & XZ^{\lceil \frac{d}{p}\rceil p}\\
X^{d-1}&X^{d-1}Z^p&\cdots&X^{d-1}Z^{\lceil \frac{d}{p}\rceil p}
\end{array}
\end{displaymath}
where $p$ is an integer which will be fixed later. Following the
well-known commutation relations of generalized Pauli group if we
ignore the phase factor, $(X^{d-1}Z^{ip})^{\dagger}Z^j$,
$(Z^j)^{\dagger}XZ^{ip}$, $(Z^{ip})^{\dagger}Z^j$,
$(Z^j)^{\dagger}Z^{ip}$ can be identified with $XZ^{-ip+j}$,
$XZ^{ip-j}$, $Z^{-ip+j}$ and $Z^{ip-j}$ respectively.

Now we divide our proof into the following two steps. First, we show
that the above gates cannot be discriminated if the ancilla system
is not allowed. If not so, then there exists a pure state
$\ket{\psi}$ which perfectly discriminates these quantum gates, and
we have the following equations:
\begin{eqnarray}
\bra{\psi}XZ^{-ip+j}\ket{\psi}&=&0\nonumber\\
\bra{\psi}XZ^{ip-j}\ket{\psi}&=&0\nonumber\\
\bra{\psi}Z^{-ip+j}\ket{\psi}&=&0\nonumber\\
\bra{\psi}Z^{ip-j}\ket{\psi}&=&0\nonumber
\end{eqnarray}
where $i$ ranges over $\{0,1,2,\cdots,\lceil\frac{d}{p}\rceil\}$ and
$j$ ranges over $\{0,1,2,\cdots,\lfloor\frac{p}{2}\rfloor\}$, but
$(i,j)\neq (0,0)$. Note that the set
\begin{displaymath}
\{XZ^{-ip+j}\}_{i,j}\bigcup\{XZ^{ip-j}\}_{i,j}\bigcup\{Z^{-ip+j}\}_{i,j}\bigcup\{Z^{ip-j}\}_{i,j}
\end{displaymath}
contains the gates $\{Z^k\}_{k=1}^{d-1}$ and $\{XZ^k\}_{k=0}^{d-1}.$
Then it holds that $\bra{\psi}XZ^k\ket{\psi}=0$ for all $0\leq k\leq
d-1$ and $\bra{\psi}Z^k\ket{\psi}=0$ for all $1\leq k\leq d-1$,
which contradicts Lemma~\ref{lemma:free}.

Second, we estimate the order of the number of the above gates. It
is obvious that the number of the gates is
$3\lceil\frac{d}{p}\rceil+3+\lfloor\frac{p}{2}\rfloor$. Then we can
choose $p=\lceil \sqrt{6d}\rceil$ and derive following inequalities
\begin{equation*}
\begin{split}
\sqrt{6d}+2 \leq  & \frac{3d}{p}+2+\frac{p}{2} \leq 3\lceil\frac{d}{p}\rceil+3+ \lfloor\frac{p}{2}\rfloor\\
\leq & \frac{3d}{p}+6+\frac{p}{2} \leq  \frac{3d}{\sqrt{6d}}+6+\frac{\sqrt{6d}+1}{2} =\sqrt{6d}+\frac{13}{2}\end{split}\end{equation*}
which completes our proof.
}

Next, we will generalize our idea to $r$-dimensional
ancilla-assisted case and show that there exist $O(\sqrt{rd})$
$d$-assisted discriminable quantum gates which cannot be perfectly
discriminated with assistance from an $r$-dimensional ancilla system.

\begin{lemma}\label{lemma:r1}
There does not exist any density operator $\rho$ with rank $\leq r$
such that $$tr(X^iZ^j\rho)=0$$ for all $1\leq i\leq r, 0\leq j\leq
d-1$, and $$tr(Z^j\rho)=0$$ for all $1\leq j\leq d-1$.
\end{lemma}
\IEEEproof{ We prove the lemma by refutation. Suppose there is a density operator $\rho$
satisfying $tr(X^iZ^j\rho)=0$ for all $1\leq i\leq r, 0\leq j\leq
d-1$, and $tr(Z^j\rho)=0$ for all $1\leq j\leq d-1$ and
$rank(\rho)\leq r$. Notice that $tr(X^iZ^j\rho)=0$ is indeed a
constraint of the $(d-i)$-th diagonal of $\rho$. We have $d$ linear
equations of the $(d-i)$-th diagonal for each $i\in \{1,2,\cdots,
r\}$. The corresponding coefficient matrix is the Vandermonde matrix
generated by $(1, \omega, \cdots, \omega^{d-1})$ which is
nonsingular. So the entries in the $(d-i)$-th diagonal of $\rho$ are
all zero for all $1\leq i\leq r$.

On the other hand, the condition that $tr(Z^j\rho)=0$ for all $1\leq
j\leq d-1$ gives us $d-1$ linear equations of the principle diagonal
of $\rho$. The coefficient matrix is a submatrix of a Vandermonde
matrix obtained by deleting the row of $(1,1,\cdots, 1)$. Thus, the
entries in the principle diagonal of $\rho$ should form a vector
which is orthogonal to the conjugate transpose of each row in the
coefficient matrix, and these entries should be the same. Since
$\rho$ is a density operator, we have $tr(\rho)=1$, which implies
that the entries in the main diagonal of $\rho$ are all nonzero.

Now consider an $(r+1)$ by $(r+1)$ submatrix $M$ of $\rho$ that lies
in the rows indexed from $1$ to ${r+1}$ and the columns indexed from
$1$ to ${r+1}$ respectively. Clearly, $M$ is upper triangular and
all the entries in its main diagonal are nonzero, and thus has
nonzero determinant. So $\rho$ has at least one nonzero
order-$(r+1)$ minor, and $rank(\rho)\geq r+1$. This contradicts our
assumption that $rank(\rho)\leq r$.}

\begin{lemma}\label{lemma:r}
There exist $O(\sqrt{rd})$ quantum gates which are $d$-assisted discriminable but are not $r$-assisted discriminable.
\end{lemma}
\IEEEproof{Let $p$ and $q$ be two positive integers which will be
specified later. We choose $\{U_i\}_{i=1}^k$ to be the following set
of gates:
\begin{displaymath}
\{X^iZ^j\}_{0\leq i\leq p-1,0\leq j\leq q-1}\bigcup \{X^{-pi'}Z^{-qj'}\}_{0\leq i'\leq \lceil\frac{r}{p}\rceil,0\leq j'\leq \lceil\frac{d}{q}\rceil} ,
\end{displaymath} where these gates are arbitrarily numbered from $U_1$
to $U_k$, and $k$ is the cardinality of the above set of gates. Then
the set $\{U_i^{\dagger}U_j\}_{1\leq i\neq j\leq k}$ contains the
following gates:
$$(X^{-pi'}Z^{-qj'})^{\dagger}X^iZ^j$$ where $0\leq i\leq p-1,$
$0\leq j\leq q-1,$ $0\leq i'\leq \lceil\frac{r}{p}\rceil,0\leq
j'\leq \lceil\frac{d}{q}\rceil$ and $(i,j,i',j')\neq (0,0,0,0)$.
Since $X$ and $Z$ are commutative, the gate
$(X^{-pi'}Z^{-qj'})^{\dagger}X^iZ^j$ can be identified with
$X^{pi'+i}Z^{qj'+j}$ if we ignore the global phase. It is easy to
see that the set $$\{X^{pi'+i}Z^{qj'+j}\}_{i,i ,j,j'},$$ with $0\leq
i\leq p-1,0\leq j\leq q-1, 0\leq i'\leq
\lceil\frac{r}{p}\rceil,0\leq j'\leq \lceil\frac{d}{q}\rceil$ and
$(i,j,i',j')\neq (0,0,0,0),$ contains all $X^iZ^j$ for $0\leq i\leq
r, 0\leq j\leq d-1$ and $(i,j)\neq (0,0)$. Thus, using
Lemma~\ref{lemma:r1} we conclude that there is not any density
operator with rank $\leq r$ satisfying $tr(X^iZ^j\rho)=0$ for all
$1\leq i\leq r, 0\leq j\leq d-1$ and $tr(Z^j\rho)=0$ for all $1\leq
k\leq d-1$. Furthermore, with Lemma~\ref{lemma:density} we see that
these quantum gates cannot be discriminated perfectly with an
$r$-dimensional ancilla system.

Now we can set $p$ and $q$ to be $\lceil\sqrt{r}\rceil$ and
$\lceil\sqrt{d}\rceil$ respectively. Then the number of quantum
gates here is
$N=pq+(\lceil\frac{r}{p}\rceil+1)(\lceil\frac{d}{q}\rceil+1)-1$
which satisfies
\begin{displaymath}
2\sqrt{rd}\leq N\leq 2(\sqrt{r}+2)(\sqrt{d}+2)-1.
\end{displaymath}
}

The upper bound given in the above lemma can be improved in some
special cases. Suppose $r\ll d$, or more specifically, $r\leq
\frac{d}{2}-1$. Then we have the following lemma which tightens our
previous upper bound significantly.
\begin{lemma}\label{lemma:rr}
Let $r\leq \frac{d}{2}-1$. Then there are $2r+2$ quantum gates which
can be $d$-assisted discriminated but cannot be $r$-assisted
discriminated.
\end{lemma}
\IEEEproof{We assume $r$ is even, let $r=2r_1$. Let $\alpha$ be the $(r+1)$-th
unity root and $\beta$ be the $(d-r-1)$-th unity root respectively.
Choose following quantum gates as $\mathcal{U}=\{U_i\}_{i=1}^k$:
\begin{eqnarray}
X_{r+1}^i\bigoplus X_{d-r-1}^i && 1\leq i\leq r_1\nonumber\\
X_{r+1}^{-i}\bigoplus \omega X_{d-r-1}^{-i} && 1\leq i\leq r_1\nonumber\\
Z_{r+1}^j\bigoplus Z_{d-r-1}^j && -r_1\leq j\leq r_1\nonumber\\
I_{r+1}\bigoplus (-1)I_{d-r-1}\nonumber
\end{eqnarray}
Here $\omega$ is some unity root. We can assume it is $p$-th unity root where $p$ is a very large prime number.

It is not difficult to check that these quantum gates can be perfectly discriminated by
$$\frac{1}{2(r+1)}I_{r+1}\bigoplus \frac{1}{2(d-r-1)}I_{d-r-1},$$ thus can be perfectly discriminated with assistance from a $d$-dimensional ancilla system.

Next, we show that there does not exist a density operator $\rho$
with $rank(\rho)\leq r$ satisfying following equations:
\begin{displaymath}
tr(U_i^{\dagger}U_j\rho)=0 \textrm{ }for \textrm{ } all \textrm{ }1\leq i\neq j\leq |\mathcal{U}|
\end{displaymath}This will be done by refutation.
Suppose there exists a density operator $\rho$ satisfying the above
equations. Then we have the following equations:
\begin{eqnarray}
tr(X_{r+1}^{-i}Z_{r+1}^j\bigoplus X_{d-r-1}^{-i}Z_{d-r-1}^j\rho)=0 \nonumber\\
\textrm{\ \ \ \ \ \ \ \ }for \textrm{ } all \textrm{ }1\leq i\leq r_1, \textrm{ }\textrm{ }-r_1\leq j\leq r_1\nonumber\\
tr(X_{r+1}^{i}Z_{r+1}^j\bigoplus \omega^{*}X_{d-r-1}^{i}Z_{d-r-1}^j\rho)=0 \nonumber\\
\textrm{\ \ \ \ \ \ \ \ }
for \textrm{ } all \textrm{ }1\leq i\leq r_1, \textrm{ } \textrm{ }-r_1\leq j\leq r_1\nonumber\\
tr(Z_{r+1}^{-j}X_{r+1}^i\bigoplus Z_{d-r-1}^{-j}X_{d-r-1}^i\rho)=0 \nonumber\\
\textrm{\ \ \ \ \ \ \ \ }for \textrm{ } all \textrm{ }1\leq i\leq r_1, \textrm{ }\textrm{ }-r_1\leq j\leq r_1\nonumber\\
tr(Z_{r+1}^{-j}X_{r+1}^{-i}\bigoplus \omega Z_{d-r-1}^{-j}X_{d-r-1}^{-i}\rho)=0 \nonumber\\
\textrm{\ \ \ \ \ \ \ \ }for \textrm{ } all \textrm{ }1\leq i\leq r_1, \textrm{ }\textrm{ }-r_1\leq j\leq r_1\nonumber
\end{eqnarray}
Let's replace the index $j$ by $-j$. Notice that
$$Z_k^{q}X_k^{p}=\omega_k^{pq}X_k^pZ_k^q$$ where $\omega_k$ is the
$k$-th unity root. Recalling that $\omega_{r+1}=\alpha$ and
$\omega_{d-r-1}=\beta$, we obtain:
\begin{eqnarray}
tr(X_{r+1}^{-i}Z_{r+1}^j\bigoplus X_{d-r-1}^{-i}Z_{d-r-1}^j\rho)=0 \nonumber\\
\textrm{ }for \textrm{ } all \textrm{ }1\leq i\leq r_1,\textrm{ }-r_1\leq j\leq r_1\nonumber\\
tr(X_{r+1}^{i}Z_{r+1}^j\bigoplus (\omega^{*})X_{d-r-1}^{i}Z_{d-r-1}^j\rho)=0 \nonumber\\
\textrm{ }for \textrm{ } all \textrm{ }1\leq i\leq r_1,\textrm{ }-r_1\leq j\leq r_1\nonumber\\
tr(\alpha^{ij}X_{r+1}^iZ_{r+1}^{j}\bigoplus \beta^{ij}X_{d-r-1}^iZ_{d-r-1}^{j}\rho)=0\nonumber\\
\textrm{ }for \textrm{ } all \textrm{ }1\leq i\leq r_1,\textrm{ }-r_1\leq j\leq r_1\nonumber\\
tr(\alpha^{-ij}X_{r+1}^{-i}Z_{r+1}^{j}\bigoplus \omega\beta^{-ij}X_{d-r-1}^{-i}Z_{d-r-1}^{j}\rho)=0 \nonumber\\
\textrm{ }for \textrm{ }all \textrm{ }1\leq i\leq r_1,\textrm{ }-r_1\leq j\leq r_1\nonumber
\end{eqnarray}
Thus, it is straightforward to derive that
\begin{eqnarray}
tr(X_{r+1}^{-i}Z_{r+1}^j\bigoplus O_{d-r-1}\rho)=0 \nonumber\\
\textrm{ } for \textrm{ }all \textrm{ }1\leq i\leq r_1, \textrm{ }\textrm{ }-r_1\leq j\leq r_1\nonumber\\
tr(X_{r+1}^{i}Z_{r+1}^j\bigoplus O_{d-r-1}\rho)=0 \nonumber\\
\textrm{ } for \textrm{ }all \textrm{ }1\leq i\leq r_1, \textrm{ }\textrm{ }-r_1\leq j\leq r_1\nonumber
\end{eqnarray}
These equations describe the constraints on the left-top
$(r+1)\times (r+1)$ submatrix $\rho_0$ of $\rho$. Furthermore,
except for the principle diagonal, each diagonal of $\rho_0$ is a
solution of these $(r+1)$ linear equations, and the coefficients of
these equations form a Vandermonde matrix generated by $(r+1)$-th
unity roots. Thus the entries in these diagonals are all zero.

Next, consider the principle diagonal of $\rho$:
\begin{eqnarray}
Z_{r+1}^j\bigoplus Z_{d-r-1}^j && -r_1\leq j\leq r_1\nonumber\\
I_{r+1}\bigoplus (-1)I_{d-r-1}\nonumber
\end{eqnarray}We have:
\begin{eqnarray}
tr(Z_{r+1}^j\bigoplus Z_{d-r-1}^j\rho)=0 && \textrm{ }for \textrm{ }all \textrm{ }-r_1\leq j(\neq 0)\leq r_1\nonumber\\
tr(Z_{r+1}^{j}\bigoplus -Z_{d-r-1}^{j}\rho)=0 && \textrm{ }for \textrm{ }all \textrm{ }-r_1\leq j(\neq 0)\leq r_1\nonumber
\end{eqnarray}
Thus,
\begin{eqnarray}
tr(Z_{r+1}^j\bigoplus O_{d-r-1}\rho)=0 && \textrm{ }for \textrm{ }all \textrm{ }-r_1\leq j(\neq 0)\leq r_1\nonumber
\end{eqnarray}
which suggests the diagonal entries in $\rho_0$ are all the same. On
the other hand,
\begin{eqnarray}
tr(I_{r+1}\bigoplus (-1)I_{d-r-1}\rho)=0\nonumber
\end{eqnarray}
implies
\begin{displaymath}
\sum\limits_{i=1}^{r+1}\rho_{ii}=\sum\limits_{i=r+2}^d\rho_{ii}.
\end{displaymath}
Since $\rho$ is positive, all entries in the principle diagonal of
$\rho$ are nonnegative. Thus, we see that $\rho_{11}$ is non zero.
Otherwise all entries in the principle diagonal of $\rho$ are all
zero, which contradicts $tr(\rho)\neq 0$.

The top-left $(r+1)\times (r+1)$ submatrix of $\rho$ is a diagonal
matrix whose principle entries are all nonzero. Thus this
$(r+1)\times (r+1)$ submatrix is nonsingular, which implies a
nonzero minor, and follows by $rank(\rho)\geq r+1$. This is a
contradiction to our assumption. Thus these quantum gates cannot be
perfectly discriminated with assistance from an $r$-dimensional
ancilla system.

For the case that $r$ is odd, above arguments still hold with a
minor modification.}

Combining Lemmas~\ref{lemma:r} and~\ref{lemma:rr}, we achieve the
main result of this subsection:
\begin{theorem}\label{theorem:upper2}
For any $r < d$, we can find $O(r)$ quantum gates which can be
$d$-assisted discriminated but cannot be $r$-assisted discriminated.
\end{theorem}
\IEEEproof{We consider the following two cases:

Case 1. $r\geq \frac{d}{2}$. From Lemma~\ref{lemma:r}, we have
$O(\sqrt{rd})$ quantum gates which can be $d$-assisted discriminated
but cannot be $r$-assisted discriminated. More specifically, for
such $r$, we can always find no more than $c\sqrt{rd}$ quantum gates
which can be $d$-assisted discriminated but cannot be $r$-assisted
discriminated where $c$ is some constant positive number. Since $d\leq 2r$,
we have $c\sqrt{rd}\leq c\sqrt{2} r$. Put $c^{\prime}=c\sqrt{2}$.
Then we can always find no more than $c^{\prime}r$ quantum gates
which can be $d$-assisted discrimination but cannot be $r$-assisted
discrimination.

Case 2. $r\leq \frac{d}{2}-1$. We only need $2r+2$ quantum gates by
Lemma~\ref{lemma:rr}.}

\subsection{Lower Bound of $r$-Reduced Problem}

We first state the main result of this subsection which presents a
lower bound for Problem 3.

\begin{theorem}\label{theorem:lower2}
Any $$k\leq \lfloor\frac{\sqrt{3(r-1)^2-19}+1}{2}\rfloor$$ quantum
gates which are $d$-assisted discriminable must also be $r$-assisted
discriminable.\end{theorem}

The proof of the above requires certain mathematical tools from
algebraic geometry. For convenience of the reader, we first recall
some definitions and results in algebraic geometry. For details, we
refer to~\cite{artin91, hartshorne77, sha95}.

Let $\mathbb{A}^n$ be an \textit{affine} $n$-space, the set of all $n$-tuples of complex numbers. We write $\mathbb{C}[x_1,x_2,\cdots,x_n]$ for the polynomial ring in $n$ variables. A
subset of $A^n$ is an \textit{algebraic set} or \textit{algebraic
variety} if it is the common zeros of a finite set of polynomials
$f_1,f_2,\cdots,f_r$ with $f_i\in \mathbb{C}[x_1,x_2,\cdots,x_n]$
for all $1\leq i\leq r$. Such an algebraic set is usually denoted by
$Z(f_1,f_2,\cdots,f_r)$. It is not hard to check that the union of a finite number
of algebraic sets is an algebraic set, and the intersection of any family of algebraic sets is again an algebraic set. Thus by taking the open subsets to be the complements
of algebraic sets, we can define a topology, called the Zariski topology on $\mathbb{A}^n$. Notice that in some references an algebraic variety should also be irreducible in the sense that it cannot be
expressed as the union of two proper algebraic sets.

Suppose we can write $\mathcal{X}=\bigcup X_i$, as a finite union of irreducible closed sets. If $X_i\nsubseteq X_j$ for all $i\neq j$, we say such a representation is irredundant, and the $X_i$ are irreducible components of $X$.

We define \textit{projective n-space}, denoted by $\mathbb{P}^n$, to be the set of
equivalence classes of $(n+1)-$tuples $(a_0,\cdots,a_n)$ of complex numbers, not all zero, under the equivalence relation given by
$(a_0,\cdots,a_n)\sim(\lambda a_0,\cdots,\lambda a_n)$ for all
$\lambda \in \mathbb{C}$, $\lambda\neq 0$.

A notion of algebraic variety may also be introduced in projective
spaces, called projective algebraic variety: a subset $Y$ of $\mathbb{P}^n$
is an \textit{algebraic set} if it is the common zeros of a finite
set of homogeneous polynomials $f_1,f_2,\cdots,f_r$ with $f_i\in
\mathbb{C}[x_0,x_1,\cdots,x_n]$ for $1\leq i\leq r$. We call open subsets of irreducible projective varieties as quasi-projective varieties.

\begin{lemma}[\cite{sha95}, Chapter 1, Section 6.2, Corollary 5]
The variety of common zeros of $r$ polynomials $f_1$, $\cdots$,
$f_r$ on an $n$-dimensional projective variety has dimension $\geq
n-r$.
\end{lemma}

\begin{theorem}[\cite{sha95}, Chapter 1, Section 6.2, Theorem 6]
Let $X, Y\subset \mathbb{P}^N$ be irreducible quasiprojective varieties with $dim(X)=n$ and $dim(Y)=m$. Then any (nonempty)
component $Z$ of $X\bigcap Y$ has $dim(Z)\geq n+m-N$.
Moreover, if $X$ and $Y$ are projective and $n+m\geq N$ then
$X\bigcap Y\neq \emptyset$.
\end{theorem}

One may notice that a matrix $M$ has rank no more than $r$ if and
only if all the determinants of its $(r+1)\times (r+1)$ submatrices
are zero. Thus all $d_A\times d_B$ matrices with rank $\leq r$ can
be considered as a variety, namely the determinantal variety, whose
projective dimension is $d_Ad_B-(d_A-r)(d_B-r)-1$. Some applications
of the determinantal variety in quantum information theory has been
found recently~\cite{Chen06, Cubitt08}. For more details, we refer
to~\cite{Harris92}.

\IEEEproof[Proof of Theorem~\ref{theorem:lower2}]{From Lemma~\ref{lemma:density}, the theorem is equivalent to the following:

\textit{Claim:} For any $$k\leq
\lfloor\frac{\sqrt{3(r-1)^2-19}+1}{2}\rfloor$$ quantum gates
$\{U_i\}_{i=1}^k$, if there is a density operator $\rho_{0}$, with
$rank(\rho_{0})=d$ and $tr(U_i^{\dagger}U_j\rho_{0})=0$ for all $
1\leq i\neq j\leq k$, then we can always find another density
operator $\rho$ satisfying $tr(U_i^{\dagger}U_j\rho)=0$ for all $
1\leq i\neq j\leq k$ and its rank is bounded by
$r$.

We will prove the above claim by applying a dimension reduction
technique:

First, the base case is $r=d$, which is quite straightforward. Since
these quantum gates are $d$-assisted discriminable, we can always
find a density operator $\rho$ satisfying these equations, thus it
is true for $r=d$.

Second, suppose for any $r$ satisfies $t\leq r\leq d$, the claim is
true, then consider the case $r=t-1$. For any given $$k\leq
\lfloor\frac{\sqrt{3(r-1)^2-19}+1}{2}\rfloor=\lfloor\frac{\sqrt{3(t-2)^2-19}+1}{2}\rfloor$$
quantum gates which can be perfectly discriminated with assistance
from a $d$-dimensional ancilla system, we will show that there
exists a density operator $\rho$ with $rank(\rho)\leq t-1$ such that
$tr(U_i^{\dagger}U_j\rho)=0$ for all $ 1\leq i\neq j\leq k$. Because
$$k\leq \lfloor\frac{\sqrt{3(t-2)^2-19}+1}{2}\rfloor\leq
\lfloor\frac{\sqrt{3(t-1)^2-19}+1}{2}\rfloor,$$ there exists a
density operator $\rho$ with $rank(\rho)\leq t$ such that
$tr(U_i^{\dagger}U_j\rho)=0$ for all $ 1\leq i\neq j\leq k$ from our
assumption. If $rank(\rho)\leq r=t-1$, then the claim is obviously
true for $r=t-1$. Otherwise, we assume $rank(\rho)=t$. Without loss
of generality, suppose
$\rho=\sum\limits_{i=1}^t\lambda_i\ket{i}\bra{i}$. Then we consider
the following two sets:
\begin{displaymath}
\mathcal{A}=\{X\in M_t|tr(U_i^{\dagger}U_jX)=0\textrm{ }\forall\textrm{ }1\leq i\neq j\leq k\}
\end{displaymath}
and
\begin{displaymath}
\mathcal{B}=\{Y \in M_t|rank(Y)\leq\lfloor\frac{t-1}{2}\rfloor\}
\end{displaymath}
Here $M_t$ is the set of $d\times d$ matrices whose entries are
nonzero only in the top left $t\times t$ submatrices. Then
$\mathcal{A}$ is a hyperplane of projective dimension at least
$t^2-1-k(k-1)$, and $\mathcal{B}$ is a determinantal variety of
projective dimension
$2t\lfloor\frac{t-1}{2}\rfloor-{\lfloor\frac{t-1}{2}\rfloor}^2-1$.
Consequently, we have:
\begin{equation*}
\begin{split}
&dim(\mathcal{A})+dim(\mathcal{B})\\
\geq &
t^2-1-k(k-1)+2t\lfloor\frac{t-1}{2}\rfloor-{\lfloor\frac{t-1}{2}\rfloor}
^2-1\\
> & t^2-1-k(k-1)-1+2t(\frac{t-3}{2})-(\frac{t-3}{2})^2-1\\
> & t^2-1.
\end{split}
\end{equation*}
So, $\mathcal{A}\bigcap\mathcal{B}\neq \emptyset$ has a component
with dimension at least $1$, we can find a nonzero matrix $H_0\in
\mathcal{A}\bigcap\mathcal{B}$. It holds that $rank(H_0)\leq
\frac{t-1}{2}$ and $tr(U_i^{\dagger}U_jH_0)=0$ for all $1\leq i\neq
j\leq k$. Notice that $tr((U_i^{\dagger}U_jH_0)^{\dagger})=0$ and
then
$$tr(U_j^{\dagger}U_iH_0^{\dagger})=tr(H_0^{\dagger}U_j^{\dagger}U_i)=0.$$
Therefore, we have $tr(U_i^{\dagger}U_jH_0^{\dagger})=0$ for all
$1\leq i\neq j\leq k$, which means $H_0^{\dagger}\in
\mathcal{A}\bigcap \mathcal{B}$, and it follows that
$H_0+H_0^{\dagger}$ and $i(H_0-H_0^{\dagger})$ cannot simultaneously
be zero matrices. Without loss of generality, suppose
$H=H_0+H_0^{\dagger}$ is not zero, we have $rank(H)\leq t-1=r$ and
$tr(U_i^{\dagger}U_jH)=0$ for all $1\leq i\neq j\leq k$. Notice the
entries in Hermitian matrix $\rho-\lambda H$ are nonzero only in the
top left $t\times t$ submatrix. Let $\rho^{\prime}$ and $H^{\prime}$
be the top-left $t\times t$ submatrices of $\rho$ and $H$
respectively. We will show that there exists $\lambda$ satisfying
$$\min\limits_{\ket{\psi^{\prime}}}\bra{\psi^{\prime}}\rho^{\prime}-\lambda
H^{\prime}\ket{\psi^{\prime}}=0,$$ which implies
$$\min\limits_{\ket{\psi}}\bra{\psi}\rho-\lambda H\ket{\psi}=0.$$
Since we can write $\rho^{\prime}=A^{\dagger}A$, where $A$ is a
nonsingular matrix, it holds that
\begin{eqnarray}
&&\min\limits_{\ket{\psi^{\prime}}}\bra{\psi^{\prime}}\rho^{\prime}-\lambda
H^{\prime}\ket{\psi^{\prime}}=0 \nonumber\\
\stackrel{\textrm{\ \ \ \ \ \ \ \ \  \ \ \ \  \ \          }}{\Longleftrightarrow} &&\min\limits_{\ket{\psi^{\prime}}}\bra{\psi^{\prime}}A^{\dagger}A-\lambda H^{\prime}\ket{\psi^{\prime}}=0\nonumber\\
\stackrel{\ket{\psi^{1}}=A\ket{\psi^{\prime}}}{\Longleftrightarrow}
&&\min\limits_{\ket{\psi^{1}}} \bra{\psi^{1}}I-\lambda
(A^{-1})^{\dagger}H^{\prime}A^{-1}\ket{\psi^{1}}=0.\nonumber
\end{eqnarray}
Since $(A^{-1)^{\dagger}}HA^{-1}$ is Hermitian, it can be expressed as
$U^{\dagger}DU$ where $U$ is unitary and $D$ is diagonal. Thus
\begin{eqnarray}
&&\min\limits_{\ket{\psi^{1}}}\bra{\psi^{1}}I-\lambda
U^{\dagger}DU\ket{\psi^{1}}=0\nonumber\\
\stackrel{\ket{\psi^{2}}=U\ket{\psi^{1}}}{\Longleftrightarrow}
&&\min\limits_{\ket{\psi^{2}}}\bra{\psi^{2}}I-\lambda
D\ket{\psi^{2}}=0.\nonumber
\end{eqnarray}
The existence of $\lambda$ that satisfies the last equation is quite
obvious now. Since $D$ is diagonal, we can express it by its entries
in the principle diagonal. Assume $D=diag\{d_1, d_2, \cdots, d_t\}$
where $d_1\geq d_2\geq \cdots \geq d_t$. If $d_1>0$, we can choose
$\lambda=\frac{1}{d_1}$ satisfying this equation. Otherwise $d_1\leq
0$ which means $d_t\leq d_{t-1}\leq \cdots\leq d_1\leq 0$, and thus
$d_t<0$. We choose $\lambda=\frac{1}{d_t}$ which also satisfies our
requirement.

Now we see that there exists $\lambda$ satisfying
$$\min\limits_{\ket{\psi^{\prime}}}\bra{\psi^{\prime}}\rho^{\prime}-\lambda
H^{\prime}\ket{\psi^{\prime}}=0,$$ and then
$$\min\limits_{\ket{\psi}}\bra{\psi}\rho-\lambda H\ket{\psi}=0.$$
So, $\rho-\lambda H$ is positive, and $rank(\rho-\lambda H)\leq
t-1$. Both $\rho$ and $H$ are in the linear subspace $\mathcal{A}$,
thus so is $\rho-\lambda H$. This implies
$tr(U_i^{\dagger}U_j\rho)=0$ for all $ 1\leq i\neq j\leq k$.
Moreover, $$\frac{\rho-\lambda H}{tr(\rho-\lambda H)}$$ is the
density operator with rank bounded by $t-1$ which can perfectly
discriminate these $\{U_i\}_{i=1}^k$. So the claim is proved for
$r=t-1$. Hence, we proved that the claim is true for any $2\leq
r\leq d$ by reduction on dimension and complete the proof of the
theorem.
}

\begin{remark}
It should be noted that above argument fails if we reduce the
dimension $r$ from 2 to 1 since
\begin{displaymath}
\mathcal{B}=\{Y \in M_r|rank(Y)\leq\lfloor\frac{t-1}{2}\rfloor=0\}
\end{displaymath}
only contains zero matrix. But this doesn't affect our argument very
much because we can choose a proper constant coefficient $c$ to make
the lower bound hold for all $r$.
\end{remark}

Theorem~\ref{theorem:lower2} can be restated as the following:
\begin{corollary}
There exists a constant positive number $c$ such that for any $k\leq cr$
quantum gates, if they are $d$-assisted discriminable, then they are
also $r$-assisted discriminable.
\end{corollary}

By comparing the above corollary with Theorem~\ref{theorem:upper2},
it can be seen that there is indeed no gap between the lower bound
and the upper bound. Thus, this corollary together with
Theorem~\ref{theorem:upper2} completely answers Problem 3.


\section{Conclusions}
In this paper, we addressed the problem of ancilla-assisted
perfectly discrimination of quantum gates. We evaluated the number
of quantum gates which can be perfectly discriminated with an
$r$-dimensional ancilla system. More specifically, we answered
following two questions.

\begin{itemize}
 \item [1] How many quantum gates  at most can be perfectly discriminated?
 \item [2] Suppose a set of quantum gates can be perfectly discriminated with some ancilla system.
 Whether their discriminability remains if the ancilla system is reduced to a small
 dimension?
\end{itemize}

For the first question, it is quite simple to observe that any
$rd+1$ quantum gates on a $d-$dimensional system cannot be perfectly
discriminated with assistance from an $r$-dimensional ancilla. And we
showed that there does exist $rd$ quantum gates on a $d-$dimensional
system which can be perfectly discriminated with assistance from an
$r-$dimensional ancilla.

Our results concerning the second question can be summarized as
follows: (1) There exists a constant positive number $c$ such that for any
$k\leq cr$ quantum gates on a $d-$dimensional system, if they are
$d$-assisted discriminable, then they are also $r$-assisted
discriminable. (2) $O(r)$ is optimal. More precisely, there are
$c^{\prime}r\textrm{ }(c^{\prime}>c)$ different quantum gates on a
$d-$dimensional system which can be $d$-assisted discriminated, but
they cannot be perfectly discriminated if the ancilla is reduced to
$r$-dimensional.

Discrimination problem is of widely interest because it has many
surprising applications. This paper represents a preliminary step
toward understanding the role ancilla system plays in discrimination
of quantum gates. There are many open questions from this approach
deserving further investigation. For example, we may combine our
model with the multi-shot discrimination or generalize the input
state to several input states as candidates.
\section*{Acknowledgment}
We are thankful to the colleagues in
the Quantum Computation and Information Research Group of Tsinghua
University for helpful discussions. And J. Chen thanks N. Yu
for bringing the original question to our attention. J. Chen also
thanks J. Yu, Z. Ji for helpful discussion. This
work was partly supported by the Natural Science Foundation of China
(Grant No. 60736011, 60621062, 60503001) and the Hi-Tech Research
and Development Program of China(863 project)(Grant No.
2006AA01Z102).

\bibliographystyle{IEEEtran}
\bibliography{IEEEabrv,power}

\end{document}